\documentclass[prx,aps,twocolumn]{revtex4-1}

\usepackage[pdftex]{graphicx}
\usepackage{amsbsy,amssymb,amsmath,bm,mathtools}

\usepackage{makecell}

\usepackage{xspace}
\usepackage{bm}% bold math
\usepackage{color}

\usepackage{url}
\usepackage[section]{placeins}
\usepackage[colorlinks=true,linkcolor=blue,citecolor=blue]{hyperref}

\begin{document}

%\preprint{Preprint}

\title{Anomalous phase separation dynamics in a correlated electron system: \\ machine-learning enabled large-scale kinetic Monte Carlo simulations}

\author{Sheng Zhang}

\author{Puhan Zhang}

\author{Gia-Wei Chern}
\affiliation{Department of Physics, University of Virginia, Charlottesville, VA 22904, USA}

\date{\today}

\begin{abstract}
Phase separation plays a central role in the emergence of novel functionalities of correlated electron materials. The structure of the mixed-phase states depends strongly on the nonequilibrium phase-separation dynamics, which has so far yet to be systematically investigated, especially on the theoretical side. With the aid of modern machine learning methods, we demonstrate the first-ever large-scale kinetic Monte Carlo simulations of the phase separation process for the Falicov-Kimball model, which is one of the canonical strongly correlated electron systems. We uncover an unusual phase-separation scenario where domain coarsening occurs simultaneously at two different scales: the growth of checkerboard clusters at smaller length scales and the expansion of super-clusters, which are aggregates of the checkerboard clusters of the same sign, at a larger scale. We show that the emergence of super-clusters is due to a hidden dynamical breaking of the sublattice symmetry. Arrested growth of the checkerboard patterns and of the super-clusters is shown to result from a correlation-induced self-trapping mechanism. Glassy behaviors similar to the one reported in this work could be generic for other correlated electron systems. 
\end{abstract}

\maketitle

Complex mesoscopic textures are ubiquitous in strongly correlated electron materials~\cite{uehara99,miao20,hanaguri04,hamidian16,moreo99,mathur03,dagotto01,emery99,dagotto_book,dagotto05,kivelson98,kivelson03,fradkin15}. Notable examples include stripe and checkerboard patterns in high-$T_c$ superconductors and nano-scale mixture of metallic and insulating domains in manganites. Not only are these mesoscopic textures of fundamental importance in correlated electron physics, they are also central to the emergence of novel functionalities in these materials. The nanoscale patterns in correlated electron materials often result from phase-separation instability driven by the electron correlation effects. Indeed, a generic feature of lightly doped Mott insulators  is the strong tendency toward phase separation in which the doped holes are expelled from the locally insulating antiferromagnetic domains~\cite{visscher74,schulz89,emery90,zaanen89,kato90,gehlhoff96,white00,yee15}. 
Although considerable efforts have been devoted to understanding the phase separation mechanisms and properties of the mixed-phase states in strongly correlated materials, the nonequilibrium pattern-formation dynamics in such systems is poorly understood.

On the other hand, intermediate states with complex structures have been observed in discontinuous phase transitions of many classical systems~\cite{gunton83,krapivsky10}. The kinetics of first-order transition is a mature subject with a long history~\cite{bray94,puri09,onuki02}. In such studies, one concerns the evolution of a system from an unstable or meta-stable state toward its preferred equilibrium phase, a process that is often characterized by the appearance of complex spatial-temporal patterns.  Several numerical techniques, ranging from kinetic Monte Carlo and molecular dynamics simulations to phase-field modeling, have been developed for large-scale simulations of phase-separation phenomena~\cite{oono88,gunton90,weinkamer04,chen02}. Of particular interest in such studies is the dynamical universality class and the associated universal growth law~\cite{hohenberg77,furukawa85}. The phase-ordering process is often modeled by partial differential equations for the order-parameter fields which describe the structure of the symmetry-breaking state. However, most of the works in this field are based on empirical energy models which often cannot capture the complex and long-ranged interaction of the order-parameter fields, especially for correlated electron systems.

A comprehensive modeling of correlation-driven phase separation thus needs to take into account the microscopic electronic processes and the mesoscopic pattern formation dynamics simultaneously. For example, one could obtain the driving forces on the order-parameter fields by integrating out the electrons on-the-fly, which means the electronic structure problem is to be solved at every time-step of the macroscopic dynamics simulation. However, the repeated solution of the electronic problem, obtained using techniques ranging from exact diagonalization to more sophisticated many-body methods, can be prohibitively expensive for large scale simulations. Such computational obstacles for multi-scale simulations are partly the reason for the lack of progress in our understanding of the phase ordering dynamics in correlated electron systems.

In this paper we make an important step towards the goal of multi-scale dynamical modeling of strongly correlated electron systems by utilizing the machine learning techniques to develop an efficient yet accurate energy model. We have thus achieved the first-ever large-scale simulation of phase separation phenomena in the Falicov-Kimball (FK) model~\cite{falicov69}, which is one of the canonical correlated electron systems.  Originally put forward as a limiting case of the Hubbard model~\cite{hubbard63}, the FK model was later independently proposed to describe semiconductor-metal transitions in rare-earth and  transition-metal compounds~\cite{falicov69}. The FK model describes conducting $c$-electrons interact with localized $f$-electrons through an on-site repulsive interaction.  Its relative simplicity allows for numerically exact solutions, which serve as important benchmarks for sophisticated many-body methods~\cite{freericks03}. The FK model itself has rich phase diagrams, and is one of the best-studied correlated electron systems that exhibit complex pattern formation and phase separation~\cite{kennedy86,freericks00,freericks02,watson95,lemanski02,tran06,maska05,maska06,antipov16}. 
In particular, FK model offers the proof of principle that stripe and checkerboard orders, which play a prominent role in the high-$T_c$ superconductivity phenomenology, can arise from pure electronic correlation effect~\cite{freericks02,freericks00,watson95,lemanski02}, such as the Kivelson-Emery scenario of phase separation.

We consider the spinless FK Hamiltonian on a square lattice~\cite{lemanski02,freericks03} in this work
 \begin{eqnarray}
 	\label{eq:H_FK}
 	\mathcal{H} = -t_{\rm nn} \sum_{\langle ij \rangle} c_i^\dagger c^{\,}_j + U \sum_i c_i^\dagger c^{\,}_i \, n^f_i.
 \end{eqnarray} 
Here $c^\dagger_i$ ($c^{\,}_i$) is the creation (annihilation) operator for a $c$-electron at site $i$, $\langle ij \rangle$ denotes nearest-neighbor pairs on the lattice, $n^f_i$  is the occupation number of the $f$-electron, $t_{\rm nn}$ is the nearest-neighbor hopping constant, which also serves as the energy unit, and $U>0$ is the strength of on-site repulsive interaction.   Thanks to the quadratic nature of the $c$-electron Hamiltonian, equilibrium phases of the FK model can in principle be exactly solved numerically by combining classical Monte Carlo method for $f$-electrons with exact diagonalization (ED) for $c$-electrons~\cite{maska05,maska06,antipov16}. Moreover, within the framework of the dynamical mean-field theory, the quantum impurity problem associated with the FK model can also be exactly solved~\cite{freericks03,tran06}.

\begin{figure}
\includegraphics[width=1.0\columnwidth]{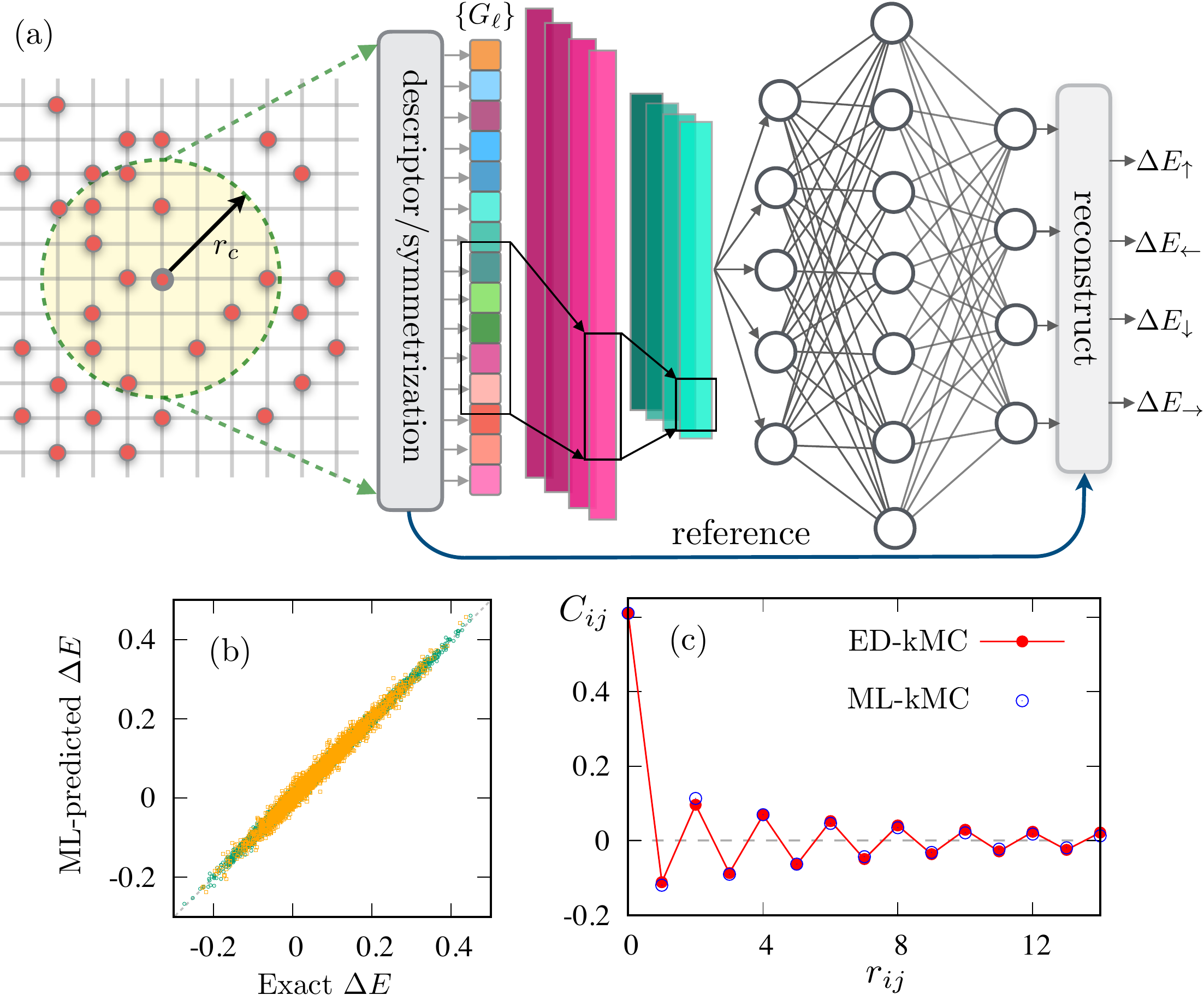}
\caption{(a) Schematic diagram of neural-network (NN) energy model for kMC dynamics simulation of the FK system. A descriptor is used to construct effective coordinates $\{G_\ell\}$  from neighborhood $f$-electron configuration $\{  n^f_j\}$ up to a cutoff~$r_c = 10$. These feature variables $\{G_\ell\}$ are input to the NN which predicts the energy differences $\Delta E$ at the output. (b) ML-predicted $\Delta E$ versus exact values; the circles and squares denote training and test datasets, respectively. (c) Comparison of $f$-electron correlation function $C_{ij} = \langle n^f_i n^f_j \rangle$ obtained from ED and ML-kMC simulations on a $30\times 30$ lattice after 5000 steps from the same initial condition.}
\label{fig:ml-scheme}  \
\end{figure}

The equilibrium phases of the square-lattice FK model have been extensively studied over the years. Exactly at half-filling for both $c$ and $f$ electrons, the ground-state exhibits a charge density wave (CDW) order with the $f$-electrons forming a checkerboard pattern~\cite{maska06,antipov16}. Away from half-filling, the model shows various stripe and incommensurate phases~\cite{lemanski02,watson95}. With slight electron or hole doping, a phase-separated regime is stabilized~\cite{lemanski02,maska05,tran06}, a scenario similar to that of the Hubbard model. Despite being one of the prominent models for electronic phase separation, the phase-ordering dynamics in FK systems has never been studied. Important questions, such as whether the system exhibits dynamical scaling and what is the domain-growth law, remain open.

To address these issues, we formulate a kinetic Monte Carlo (kMC) method~\cite{weinkamer04,barkema09} to simulate the phase ordering process of the FK model subject to a temperature quench. While the $c$-electrons have well-defined dynamics in the FK Hamiltonian, the $f$-electrons in the FK model, with $n^f_i = 0$ or 1,  are static discrete variables, similar to classical Ising spins. To endow the $f$-electron with a dynamics, a random-walk algorithm is used to model their diffusive motion. At every time-step, an attempt is made to move a randomly chosen $f$-electron to one of its empty neighbors. Whether the update is accepted is determined by the standard Metropolis criterion~\cite{weinkamer04}. We further assume that the equilibration of $c$-electrons is much faster compared with the random walks of $f$-electrons, analogous to the Born-Oppenheimer approximation in quantum molecular dynamics~\cite{marx09}. Consequently, the motion of the heavier $f$-electrons depends on the free-energy of the quasi-equilibrium $c$-electrons before and after the update. The acceptance probability of such a nearest-neighbor move is 
\begin{eqnarray}
	p_{i\to j} = \frac{1}{4} {\rm min}\left(1, e^{-\Delta E_{i\to j}/k_B T}\right), 
\end{eqnarray}
where $\Delta E_{i\to j}$ is the free-energy difference of $c$-electrons due to the hopping of $f$-electron from site $i$ to $j$. The probability that the $f$-electron stays put is $p_{i\to i} = 1 - \sum_{j} p_{i \to j}$.

The $c$-electron free-energy can be computed using either ED or more efficient techniques such as kernel polynomial method~\cite{weisse06,furukawa04,alvarez05,wang18}. However, the quantum kMC simulation described above is very time-consuming for large systems, since one needs to solve the electron tight-binding problem four times at every time-step in order to update just one $f$-electron. To overcome this computational bottleneck, we apply the machine learning (ML) methods that have been exploited to improve the efficiency of quantum molecular dynamics simulations~\cite{behler07,bartok10,zhang18,suwa19,noe20,li15,smith17}. Similar approaches have also been used to enable large-scale quantum spin dynamics in double-exchange systems~\cite{zhang20,zhang21}. The central idea of our approach, summarized in Fig.~\ref{fig:ml-scheme}(a), is based on the principle of locality~\cite{kohn96,prodan05}, or what W. Kohn termed the nearsightedness of electronic matter. In our case, the locality principle indicates that the energy change $\Delta E_{i \to j}$ depends only on $f$-electron configuration in the neighborhood of the local update. Specifically, the energy change $\Delta E_{i \to j}$ of a local update is assumed to depend on neighborhood configuration through universal functions:
\begin{eqnarray}
	\Delta E_{i \to j} = \varepsilon(\hat{\mathbf n}_{ij}, \mathcal{C}_i),
\end{eqnarray}
where $\hat{\mathbf n}_{ij} = \pm \hat{\mathbf x}, \pm \hat{\mathbf y}$ denotes the orientation of the $i\to j$ bond, and $\mathcal{C}_i = \bigl\{ n^f_j \, \big| \, |\mathbf r_j - \mathbf r_i | \le R_{\rm cutoff} \bigr\}$ describes the neighborhood $f$-electron configuration up to a cutoff radius $R_{\rm cutoff}$. The complex dependence of the energy function $\varepsilon(\cdot)$ on the local environment is then encoded in a neural network to be trained by exact solutions from small systems.

Next, the effective energy function $\varepsilon(\cdot)$ is expected to also preserve the site-symmetry of the lattice. To this end, we first note that the distribution $\{ n^f_j \}$ of $f$-electrons in the neighborhood corresponds to a reducible representation of the point group associated with site-$i$. By decomposing it into the irreducible representations (IRs), the neighborhood configuration~$\mathcal{C}_i$ is translated into the amplitudes of the IRs~\cite{ma19,kondor07,bartok13}. Effective coordinates $\{G_\ell \}$ that are invariant under the symmetry operations of the on-site point group are obtained from the bispectrum coefficients of the IRs. Details of this construction can be found in the  Appendix~\ref{sec:descriptor}.

The generalized coordinates $\{G_\ell\}$, also known as feature variables, are then fed into a neural network (NN). Due to the discrete binary nature of $f$-electron occupation number, a convolutional NN is used to enhance recognition of the major features in the input. The output from the convolutional layers then propagates to a fully connected feed-forward NN, which in turn produces the predicted energy differences $\Delta E_{i\to j}$. We have built an eight-layer NN model trained by dataset obtained from ED-kMC simulations on a 30$\times 30$ lattice; see the Appendix~\ref{sec:NN} for details of the neural network structure, dataset selection, and training process. As shown in Fig.~\ref{fig:ml-scheme}(b), the ML-predicted energy difference $\Delta E_{i \to j}$ agree well with the exact values. We further show that $f$-electron correlation function $C_{ij} = \langle n^f_i n^f_j \rangle$ obtained from kMC simulations based on the NN-model also agrees well with that of exact kMC simulations; see Fig.~\ref{fig:ml-scheme}(c).

\begin{figure}[t]
\includegraphics[width=0.99\columnwidth]{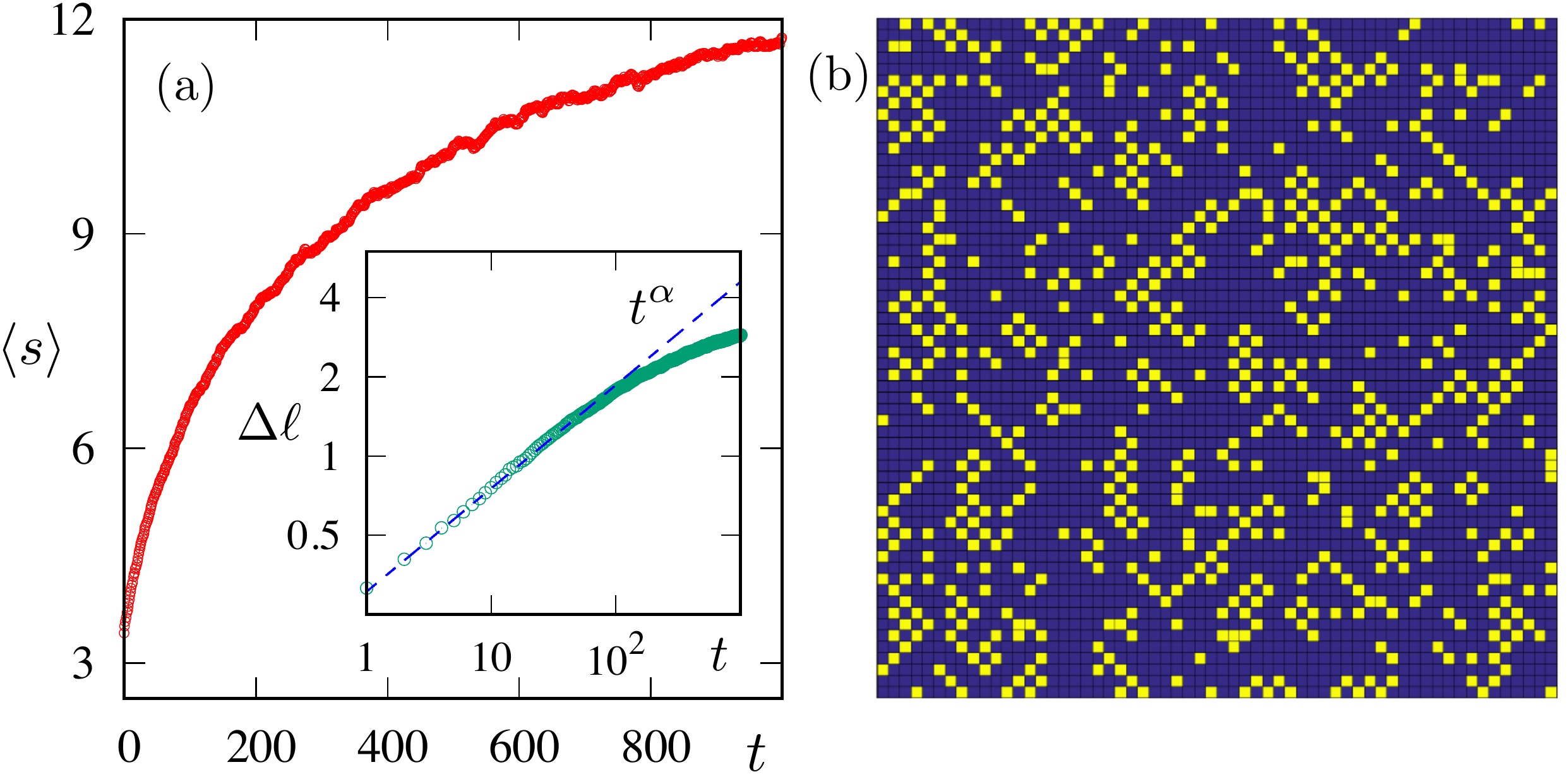}
\caption{(Color online) (a) Average size $\langle s \rangle$ of checkerboard cluster as a function of time obtained from ML-kMC on a $150\times 150$ lattice. The inset shows the time dependence of the characteristic length scale $\ell(t) = \ell_0 + \Delta \ell$, where $\ell = \langle s\rangle^{1/2}$. The dashed line indicates a power-law growth with exponent $\alpha \approx 0.35$. Here time is measured in terms of 100 MC steps. (b)~A close-up view of $f$-electron configuration at $t =800$ after quench.  }
\label{fig:cdw-cls}  \
\end{figure}

With the properly benchmarked NN energy model, we perform large-scale ML-kMC simulations on the FK model with up to $10^5$ lattice sites. Our goal is to study the growth dynamics of checkerboard clusters after a temperature quench. To this end, we consider slightly doped $c$-electrons with a filling fraction $\rho_c = 0.55$, and a low $f$-electron density of $\rho_f = 0.187$. The repulsive interaction is set at~$U = 2\, t_{\rm nn}$. The low temperature phase corresponding to these parameters is a phase-separated mixture of the checkerboard CDW ordering of $f$-electrons, and the metallic region in the absence of $f$-electrons~\cite{tran06}. Some stripe orders have also been observed.  In our simulations, the system is initially prepared in a state with random distribution, and the temperature is suddenly reduced to~$T = 0.05\, t_{\rm nn}$ at time $t = 0$. 
A snapshot of the $f$-electron configuration at a late time after quench, shown in Fig.~\ref{fig:cdw-cls}(b), clearly displays several checkerboard clusters and some diagonal stripes of the $f$-electrons.

Fig.~\ref{fig:cdw-cls}(a) shows the increase of average checkerboard cluster size $\langle s \rangle$ with time, indicating the aggregation of $f$-electrons to form CDW order during the relaxation.  Since the number of $f$-electrons is conserved, the growth of checkerboard domains is similar to the phase separation of a conserved order parameter which is expected to follow the $t^{1/3}$ power-law growth  predicted by the Lifshitz-Slyozov-Wagner (LSW) theory~\cite{lifshitz61,wagner61}, or the model-B dynamical model~\cite{hohenberg77,cahn58}. As shown in the inset of Fig.~\ref{fig:cdw-cls}(a), the typical length scale of checkerboard clusters indeed increases according to a power-law $\Delta\ell   \sim  t^{\alpha}$, even early in the phase-separation process, although the exponent $\alpha \approx 0.35$ is slightly higher than the LSW prediction. The discrepancy here can be attributed to the fact that the LSW scaling only occurs for coarsening of very large domains at late times~\cite{huse86}.

\begin{figure*}[t]
\includegraphics[width=1.95\columnwidth]{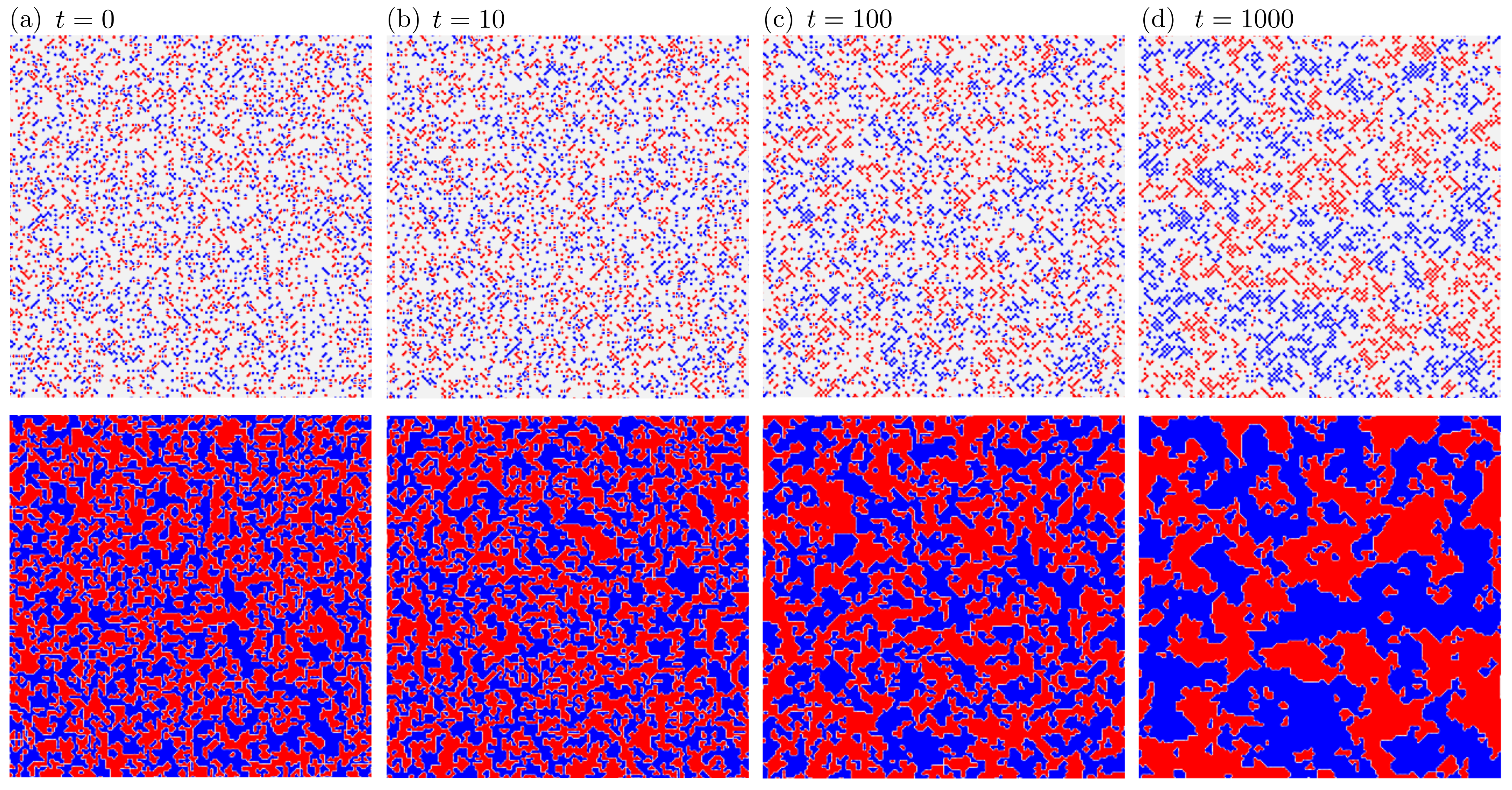}
\caption{(Color online) Top: snapshots of $f$-electron configuration obtained from kMC simulations of phase separation in a $150\times 150$ square-lattice FK model. The blue and red dots indicate $f$-electrons on the A- and B-sublattice, respectively. Bottom: configurations of Ising variables $\sigma_i$ that characterize the  $Z_2$-symmetry-breaking domains associated with super-clusters. 
}
\label{fig:snapshots}  \
\end{figure*}

\begin{figure}[t]
\includegraphics[width=1.0\columnwidth]{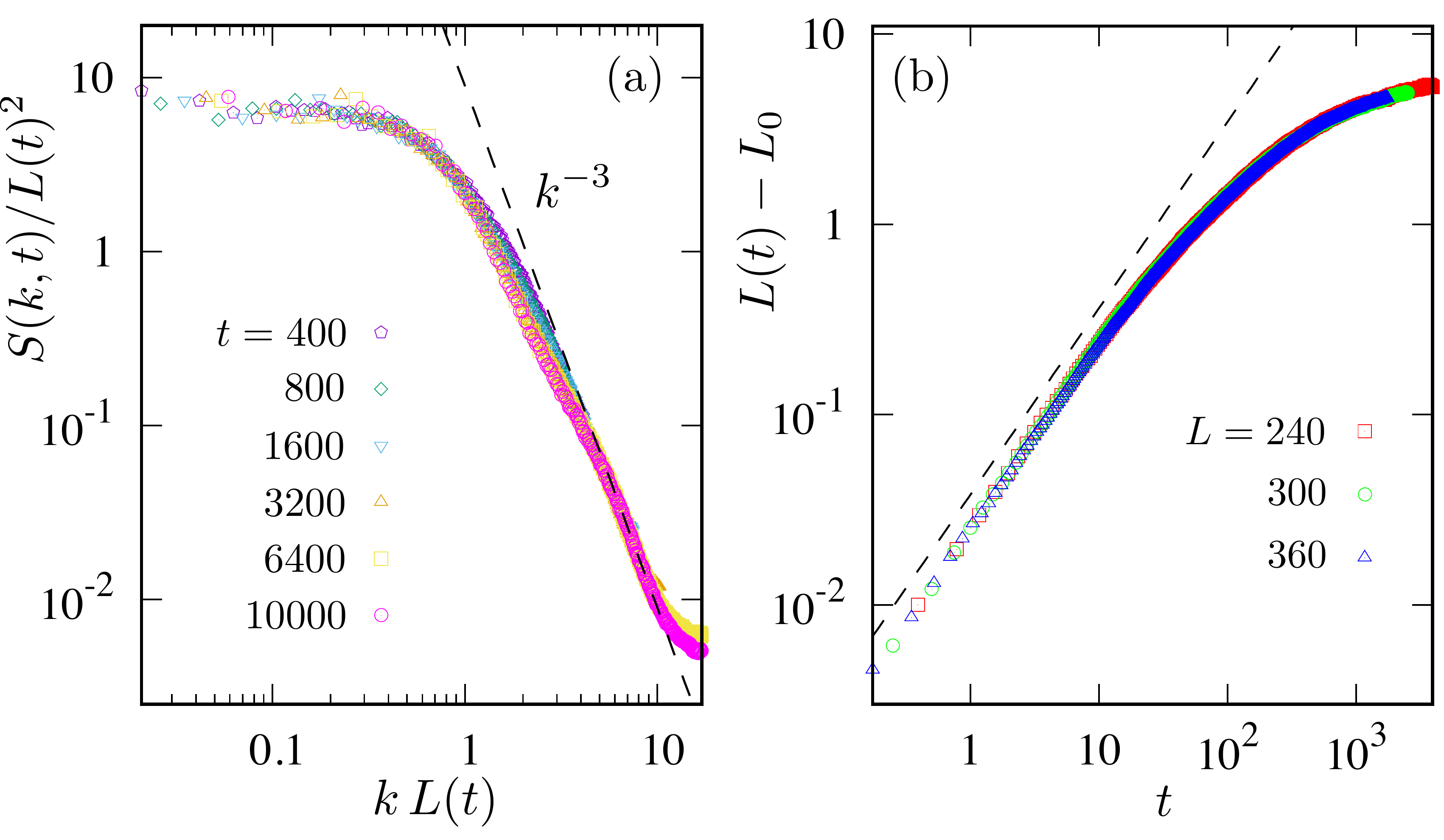}
\caption{(Color online) (a) Scaling plot of the time-dependent structure factor $S(\mathbf k, t)$ obtained from Fourier transform of the $Z_2$ order parameter. The dashed line shows the $k^{-3}$ Porod's law in 2D. (b) Characteristic length $L(t)$ of the super-clusters as a function of time for three different lattice sizes.  The dashed line indicates the linear growth $\Delta L(t) \sim t$.}
\label{fig:scaling}  \
\end{figure}

However, this power-law regime only lasts a short duration and the growth slows down significantly at late stage. This stagnation of domain-growth  cannot be attributed to the finite size effect since the average cluster size is still significantly smaller than the system sizes at late times.  Instead, the freezing of the checkerboard clusters, which will be discussed in more detail below, is related to a sublattice symmetry breaking hidden in the phase separation process. To illustrate this effect, we use different colors to label the $f$-electrons at the two sublattices of the square lattice. As shown in the top row of Fig.~\ref{fig:scaling}(a)--(c), while the sizes of checkerboard clusters remain small, $f$-electrons residing on the same sublattice tend to stick together, thus forming super-clusters of the checkerboards. Crucially, the formation of such super-structures also breaks the $Z_2$ sublattice symmetry.

It is thus worth noting that there are two $Z_2$ symmetry breaking taking place simultaneously during the nonequilibrium relaxation. First, the formation of a checkerboard cluster itself spontaneously breaks the sublattice symmetry of a bipartite lattice at a smaller length scale: the CDW order has to reside on either the A or B-sublattice. We can characterize this symmetry-breaking by assigning a polarity $+1$ ($-1$) to a checkerboard if the $f$-electrons are on the A (B) sublattice.  The correlation length of this smaller-scale $Z_2$ symmetry breaking is determined by the characteristic sizes of the CDW clusters, which remain small (less than 10 lattice constants). On the other hand, as discussed above, there is another hidden $Z_2$ sublattice symmetry breaking occurring at larger length scales, which corresponds to the  emergence of the super-clusters, i.e. clusters of CDW's. The super-cluster can be viewed as the aggregation of disconnected checkerboards of the same polarity.
 
To describe the larger-scale $Z_2$ symmetry-breaking associated with the coarsening of super-clusters, we introduce an Ising variable $\sigma_i$ at every lattice site such that $\sigma_i = +1$~($-1$) if the $f$-electron closest to site-$i$ belongs to the A(B) sublattice. Based on this definition,  the bottom row of Figs.~\ref{fig:scaling}(a)--(c) shows the three Ising configurations $\{\sigma_i\}$ that correspond to the respective $f$-electron distribution $\{n^f_i\}$  shown on the top. In terms of the Ising spin language, the clustering of checkerboards into super-clusters thus corresponds to the growth of Ising ferromagnetic domains. The order parameter $\phi$ describing this $Z_2$ symmetry-breaking is then given by the magnetization density of Ising spins, i.e. $\phi = \langle \sigma_i \rangle$. It is worth noting that $\phi$ is not conserved in the kMC dynamics of $f$-electrons. Phenomenologically, such non-conserved field is governed by the time-dependent Ginzburg-Landau equation (TDGL) or model-A dynamics~\cite{hohenberg77}. The resultant domain-coarsening is characterized by the Allen-Cahn power law~\cite{bray94,onuki02}
\begin{eqnarray}
	L \sim t^{1/2}.
\end{eqnarray} 
However, as we will show next, the coarsening of super-clusters in our case does not follow the expected power-law due to an unusual self-confinement of the $f$-electrons.

To characterize the growth of Ising domains associated with the super-clusters, we  compute the structure factor of the Ising spins: $S(\mathbf k, t) = \bigl| \frac{1}{N} \sum_i \sigma_i(t) \exp(i \mathbf k \cdot \mathbf r_i) \bigr|^2$. The ferromagnetic ordering implies that $S(\mathbf k, t)$ exhibits a growing peak at $\mathbf k = 0$. The inverse of the width of this peak can be used as a measure of the characteristic length scale of the super-clusters: $L^{-1}(t) \sim \Delta k  =  \sum_{\mathbf k} S(\mathbf k, t) |\mathbf k| / \sum_{\mathbf k} S(\mathbf k, t) $. Using this characteristic length as a scale factor, Fig.~\ref{fig:scaling}(d) shows the scaled time-dependent structure factor versus the dimensionless momentum $|\mathbf k| L(t)$. As can be seen from the figure, the data points at different times collapse roughly on the same curve, indicating that the coarsening of the Ising domains exhibits a dynamical scaling, 
\begin{eqnarray}
	S(\mathbf k, t) = L^2(t) \mathcal{G}[ |\mathbf k| L(t) ], 
\end{eqnarray}
where $\mathcal{G}(x)$ is a universal scaling function. The $1/k^{3}$ power-law tail at large momenta, consistent with the 2D Porod's law~\cite{puri09}, results from the sharp interfaces between the two different types of Ising domains, or super-clusters with opposite polarities.

The characteristic length $L(t)$ extracted from the structure factor is shown in Fig.~\ref{fig:scaling}(e) as a function of time for three different lattice sizes. Interestingly, except for a short initial period (up to $t \sim 10$), the growth of this length scale does not follow the expected power law, especially at late times. Moreover,  even the initial seemingly power-law growth is not consistent with the $\alpha = 1/2$ Allen-Cahn law. Instead, $L$~seems to increase linearly with time in this initial regime. To understand this anomalous behavior, we note that the TDGL equation or the Allen-Cahn theory describes an interface-controlled domain growth where the interfacial velocity is proportional to the curvature of the domain-interface~\cite{allen79}. On the other hand, since the $Z_2$ order parameter in our case is defined by whether the aggregating $f$-electrons are on the A- or B-sublattice, the resultant domain growth needs not rely on the expansion of an existing boundary. Instead, a super-cluster can quickly increase its size as $f$-electrons in its neighborhood move from one sublattice to another via only a nearest-neighbor hopping. Due to such collective movement of $f$-electrons, the growth of the super-clusters exhibits an avalanche-like behavior similar to the Barkhausen effect in magnetic domain growth.  A faster linear growth of the super-clusters thus arises from such avalanche dynamics at the early stage. As will be discussed below, such collective behavior is induced by an effective non-local interaction between $f$-electrons.

\begin{figure}
\includegraphics[width=0.99\columnwidth]{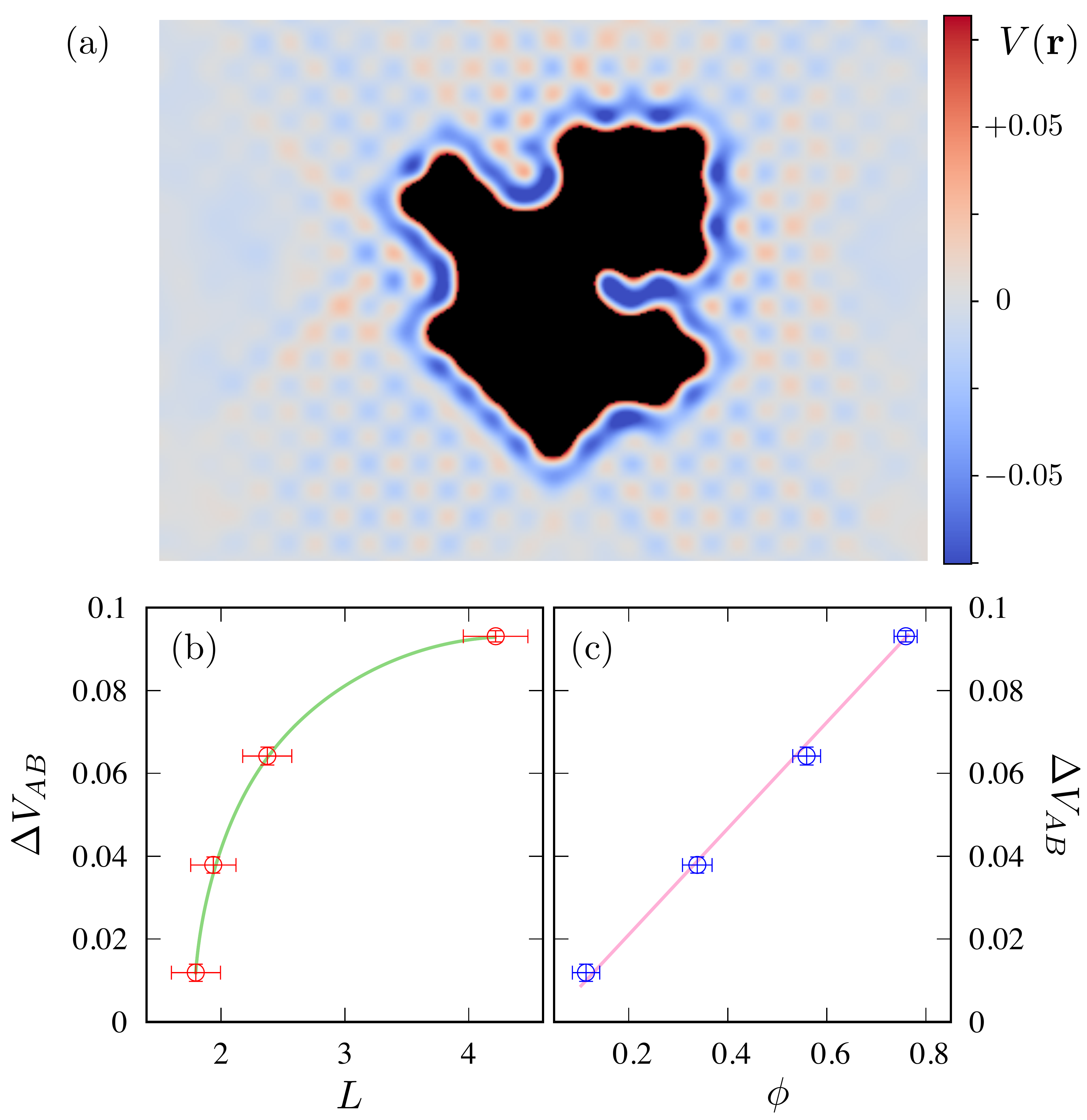}
\caption{(Color online) (a) Density plot of effective potential $V(\mathbf r_i)= -k_B T \log \nu_i$ for $f$-electrons created by a checkerboard cluster at the center. The potential field $V(\mathbf r)$ exhibits the same staggering patten as that of the checkerboard cluster at the center. Ions in the neighborhood thus tend to reside on the same sublattice, leading to the growth of the super-cluster. The depth of the staggering potential, given by the averaged potential difference between the two sublattices $\Delta V_{AB}$ versus (b) the characteristic length $L$ of super-clusters and (c) the Ising order parameter $\phi$.}
\label{fig:stagger-v}  \
\end{figure}

Although the repulsion $U$ between the two types of electrons is local in the FK model, the heavier $f$-electrons experience an effective long-range interaction mediated by the itinerant $c$-electrons. In particular, due to this non-local interaction, the presence of a checkerboard cluster creates a staggered potential  in its neighborhood that takes alternating values on neighboring sites of the bipartite lattice.  This effective potential is illustrated in Fig.~\ref{fig:stagger-v}(a) where a test $f$-electron is placed in the neighborhood of a checkerboard cluster at the center. Exact MC simulation was used to obtain the frequency $\nu_i$ that the test particle  stays at site $i$, from which the potential is computed: $V_i = -k_B T \log \nu_i$. As shown in Fig.~\ref{fig:stagger-v}(a), the effective potential field $V(\mathbf r_i)$ exhibits the same staggered pattern whose polarity is determined by that of the center checkerboard cluster. Consequently, $f$-electrons in the neighborhood of this checkerboard cluster tend to be trapped in the same sublattice. Also importantly, this staggered potential causes existing checkerboards in the neighborhood to change its polarity, thus leading to the formation of a super-cluster and its subsequent growth that is not captured by the interface-controlled mechanism.

At late stage of the phase separation, a much slower logarithmic-like growth sets in for super-clusters, as shown in Fig.~\ref{fig:scaling}(e). Interestingly, exactly the same staggered potential discussed above is also responsible for the stalled growth of the Ising domains and, in fact, of the smaller checkerboard clusters as well. To understand this unusual freezing behavior, we note that while the strength of the staggered potential  increases with the size of CDW cluster from which it originates, the energy barrier $\Delta V_{AB}$ between the two sublattices is also enhanced as  more and more checkerboards merge to form a larger super-cluster. To demonstrate this effect, we consider different geometrical arrangements of a finite number of checkerboard clusters on a $30\times 30$ lattice, giving rise to different shapes and sizes of super-clusters or Ising domains. MC simulation with exact diagonalization is then used to compute the resultant effect potential $V(\mathbf r_i)$ for the $f$-electrons; details can be found in Appendix~\ref{sec:stagger-v}. The average potential difference between the A/B-sublattices 
\begin{eqnarray}
	\Delta V_{AB} = \Bigl\langle \frac{1}{N/2} \Bigl| \sum_{i\in {\rm A}} V_i - \sum_{i \in {\rm B}} V_i \Bigr| \Bigr\rangle,
\end{eqnarray}  
is shown in Fig.~\ref{fig:stagger-v}(b) as a function of the numerically obtained characteristic length $L$ of the Ising domains. The staggered potential $\Delta V_{AB}$ indeed increases with the size of the super-clusters. Moreover, we observe an intriguing linear dependence of the potential barrier $\Delta V_{AB}$ on the effective Ising order parameter $\phi$, as shown in Fig.~\ref{fig:stagger-v}(c).

Importantly, the fact that the energy barrier $\Delta V_{AB}$  increases with the size of the Ising domains also explains the freezing behaviors observed in our ML-kMC simulations. As the size $L$ of super-clusters increases with time, the potential difference between the two sublattices  becomes so strong that individual $f$-electrons are deeply trapped at one sublattice and cannot hop to the neighboring sites. For example, consider a checkerboard cluster on sublattice-A   in Fig.~\ref{fig:stagger-v}(a) and a test particle sitting at a site that belongs to the lower-energy sublattice-A. Although the checkerboard at the center has a strong affinity to the new particle, as evidenced by the rather low potential energy at the edge of the cluster, the large energy barrier at B-sublattice prevents the $f$-electron from joining the cluster. The reduced mobility of the $f$-electron thus results in an arrested coarsening of both the super-clusters as well as the smaller-sized checkerboard clusters.

To summarize, by utilizing modern machine learning techniques, we have successfully developed a neural network energy model that allows us to perform the first-ever large-scale kinetic Monte Carlo simulation on the well-studied FK~model. We discover a novel phase-ordering phenomenon where domain coarsening occurs simultaneously at two different scales: the growth of checkerboard clusters and the expansion of Ising domains associated with a hidden broken sublattice symmetry. The competition of these two processes leads to an anomalously slow phase separation. Several interesting dynamical phenomena, such as the early-stage avalanche domain growth and the decelerated coarsening of super clusters, require further investigation and will be left for future work. 

Unusual domain-coarsening has been reported in classical systems, which is often related to frustrated interactions or quenched disorder~\cite{shore92,evans02,tanaka00,zannetti09,corberi15}. In this work we describe a new freezing mechanism which arises from the interaction of itinerant $c$-electrons and classical $f$-electrons. Similar glassy dynamics could be generic for phase ordering in other correlated electron systems. A characteristic feature of correlated electron materials is the coexistence of fast electron quasiparticles and slow bosonic or collective degrees of freedom.  The nontrivial interplay between these two sets of variables could lead to novel dynamical phenomena that are unique to correlated electrons. Given the complexity of such systems, we envision ML techniques as an indispensable tool for multi-scale modeling of nonequilibrium dynamics driven by electron correlation effect.

\bigskip

\begin{acknowledgements}
The work was supported by the US Department of Energy Basic Energy Sciences under Contract No. DE-SC0020330. The authors acknowledge Research Computing at The University of Virginia for providing computational resources and technical support that have contributed to the results reported within this publication.
\end{acknowledgements}

\appendix

\section{Descriptor}

\label{sec:descriptor}

As described in the schematic diagram for the neural-network (NN) model in the main text, a descriptor is used to construct effective coordinates $\{G_\ell\}$ from neighborhood $f$-electron configuration $\{n^f_j\}$ up to a cutoff radius $R_{\rm cutoff} = 10$. These feature variables $\{ G_\ell \}$ are then fed into the NN which predicts the energy differences $\Delta E_{i \to j}$ caused by the update at the output. As the energy prediction should preserve the discrete lattice symmetry of the square-lattice Falicov-Kimball (FK) model, the generalized coordinates $\{ G_\ell \}$ are expected to be invariant under symmetry operations of the discrete point group, for example 90$^\circ$ rotation about the $z$-axis, or mirror reflection about $xz$, $yz$ planes.   To this end, we first use the fact that the $f$-electron occupation-numbers $\{n^f_j\}$ within the cutoff form the basis of a high-dimensional representation of the point group, which in the case of square-lattice is the D$_4$ group. By decomposing this high-dimensional representation into the irreducible representations (irrep) of the symmetry group, invariants can be systematically constructed from the basis functions of the various irreps~\cite{grouptheory}.

\begin{figure}[b]
     \centering
     \includegraphics[width=0.22\textwidth]{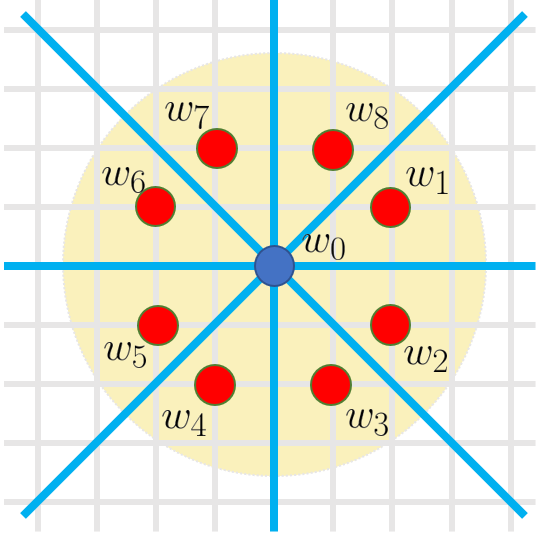}
    \caption{The example of an 8-dimensional block of the $f$-electron configuration $\{w_j\}$ corresponding to the fourth nearest neighbors of the central $f$-electron at site-0.  }
    \label{fig1}
\end{figure}

\begin{figure*}
   \centering
    \includegraphics[width=\textwidth]{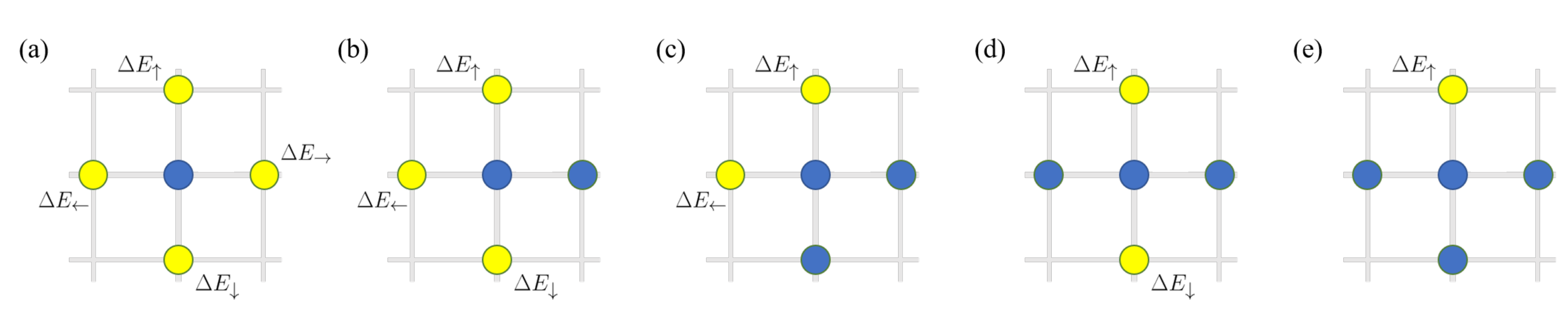}
    \caption{Five different nearest-neighbor configurations for updating an $f$-electron at the central site $n^f_0 = 1$. The blue circles denote nearest neighbor sites that are occupied by an $f$-electron, while the yellow circles denote empty sites. }
    \label{fig2}
\end{figure*}

Since the distance is preserved by symmetry operations of the point group, the representation matrices of the neighborhood occupation are automatically block-diagonalized according to the distances to the central point. This significantly simplifies the task of finding the irreps. For example, Fig.~\ref{fig1} shows the case of a neighboring sites forming an 8-dimensional block, which can be readily decomposed as $8 = A_1\oplus A_2\oplus B_1\oplus B_2\oplus 2E$.   The corresponding basis of the irreps are
\begin{eqnarray*}
        f^{A_1}&=&w_1+w_2+w_3+w_4+w_5+w_6+w_7+w_8 \\
        f^{A_2}&=&w_1-w_2+w_3-w_4+w_5-w_6+w_7-w_8 \\
        f^{A_3}&=&w_1+w_2-w_3-w_4+w_5+w_6-w_7-w_8 \\
        f^{A_4}&=&w_1-w_2-w_3+w_4+w_5-w_6-w_7+w_8 \\  
        \bm{f}^{E_1}&=&(w_2-w_3-w_6+w_7,\ w_1+w_4-w_5-w_8) \\ 
        \bm{f}^{E_2}&=&(w_1+w_2-w_5-w_6,\ w_3+w_4-w_7-w_8) 
\end{eqnarray*}
By repeating the same procedures for each block, we arrange the resultant irrep basis functions into a vector $\bm{f}^\Gamma_r=(f^\Gamma_{r,1},f^\Gamma_{r,2},\cdots,f^\Gamma_{r,D_{\Gamma}})$ where $\Gamma$ labels the irrep, $r$ enumerates the multiple occurrence of irrep $\Gamma$ in the decomposition of the $f$-electron configuration, and $D_{\Gamma}$ is the dimension of the irrep. Given these basis functions, one can immediately obtain a set of invariants called power spectrum $\{p^\Gamma_r\}$, which are the amplitude of each individual irrep functions, i.e. $p^\Gamma_r = \left| \bm f^\Gamma_r  \right|^2$. However, feature variables based only on power spectrum are incomplete in the sense that the relative phases between different irreps are ignored. For example, the relative ``angle" between two irreps of the same type: $\cos\theta = (\bm{f}^{\Gamma}_{r_1}\cdot\bm{f}^{\Gamma}_{r_2})/|\bm{f}^{\Gamma}_{r_1}||\bm{f}^{\Gamma}_{r_2}|$ is also an invariant of the symmetry group. Without such phase information, the NN model might suffer from additional error due to the spurious symmetry, namely two irreps can freely rotate independent of each other.

A systematic approach to include all relevant invariants, including both amplitudes and relative phases, is the bispectrum method~\cite{kondor07,bartok13}. In this work, we develop a descriptor similar to the one used in Ref.~\cite{ma19}, that is modified from the bispectrum method. We introduce a set of reference basis functions $\bm f^{\Gamma}_{\rm ref}$ for each irrep of the point group. These reference basis are computed by averaging large blocks of bond and chirality variables, such that they are less sensitive to small changes in the neighborhood spin configurations. We then define the relative ``phase" of a irrep as the projection of its basis functions onto the reference basis: $\eta^\Gamma_r \equiv \bm f^\Gamma_r \cdot \bm f^\Gamma_{\rm ref} / |\bm f^\Gamma_r |\, |\bm f^\Gamma_{\rm ref}|$. The effective coordinates are then obtained from the  power spectrum and the relative phases: $\{ G_\ell \} = \{ p^{\Gamma}_r \,\, , \,\, \eta^\Gamma_r \}$.

It is worth noting that the energy cost of hopping to a neighbor that is already occupied by another $f$-electron is infinite. Numerically, It is very difficult to include this special situation among finite predictions. However, for such infinite energy cost, the corresponding probability is zero, which means we can preclude such transition probability in our consideration for such situations.  Consequently, it is easier in practice to implement several different NN models, one for each of the five different nearest-neighbor $f$-electron configurations shown in Fig.~\ref{fig2}. Importantly, depending on the nearest-neighbor configuration, different point-group symmetry has to be used for computing the generalized coordinates $\{ G_\ell \}$. Our discussion of the descriptor above is mainly focused  on the case of four empty neighbors shown in Fig.~\ref{fig2}(a). For configurations shown in Fig.~\ref{fig2}(b), (c), and (e), the symmetry group is reduced to C$_2$, while the situation shown in Fig.~\ref{fig2}(d) is described by point group D$_2$.

\section{Neural network model And benchmark}

\label{sec:NN}

\begin{table}%The best place to locate the table environment is directly after its first reference in text
\begin{ruledtabular}
\begin{tabular}{|c|cc|}
\textrm{Layer}&\textrm{Network}&\\
\colrule
Input Layer & [316,1]\footnote{The shape of the input data [one dimensional dataset length, No. channels]}& \\
\hline
Convolutional Layer 1 & \makecell[c]{conv(5,1,0,16)\footnote{one-dimensional convolutional filter with arguments (filter size, stride, padding, No. filters).}\\maxpool(3,3)\footnote{ Max-pooling layer with arguments (pool size, stride length).}\\act\footnote{The activation function.} =ReLU} &\\
\hline
Convolutional Layer 2 & \makecell[c]{conv$(5,1,1,32)$\\maxpool$(3,3)$\\act=ReLU\\flatten\footnote{flatten the multi-channel output of the previous layer to the one-dimensional neurons.} $\rightarrow$[1088]} &\\
\hline
Hidden Layer 3 & \makecell[c]{FC(1088,256)\footnote{Fully connected layer with arguments (input size, output size).}\\act=ReLU} &\\
\hline
Hidden Layer 4 & \makecell[c]{FC(256,128)\\act=ReLU} &\\
\hline
Hidden Layer 5 & \makecell[c]{FC(128,64)\\act=ReLU} &\\
\hline
Hidden Layer 6 & \makecell[c]{FC(64,32)\\act=ReLU} & \\
\hline
Output Layer & FC(32,1) & \\
\end{tabular}
\end{ruledtabular}
\label{table1}
\caption{The NN structure and parameters}
\end{table}

As discussed in the main text, we have built an 8-layer neural network model on PyTorch~\cite{pytorch} for the large-scale ML-kMC simulations of the FK model. Details of the NN model  are summarized in Table~\ref{table1}. We have included convolutional layers in our NN to extract characteristic patterns of the input $f$-electron configurations~\cite{cnn}. Notably, we have verified that NN models with additional convolutional layers outperform those based only on fully connected layers.

\begin{figure}
     \centering
     \includegraphics[width=0.45\textwidth]{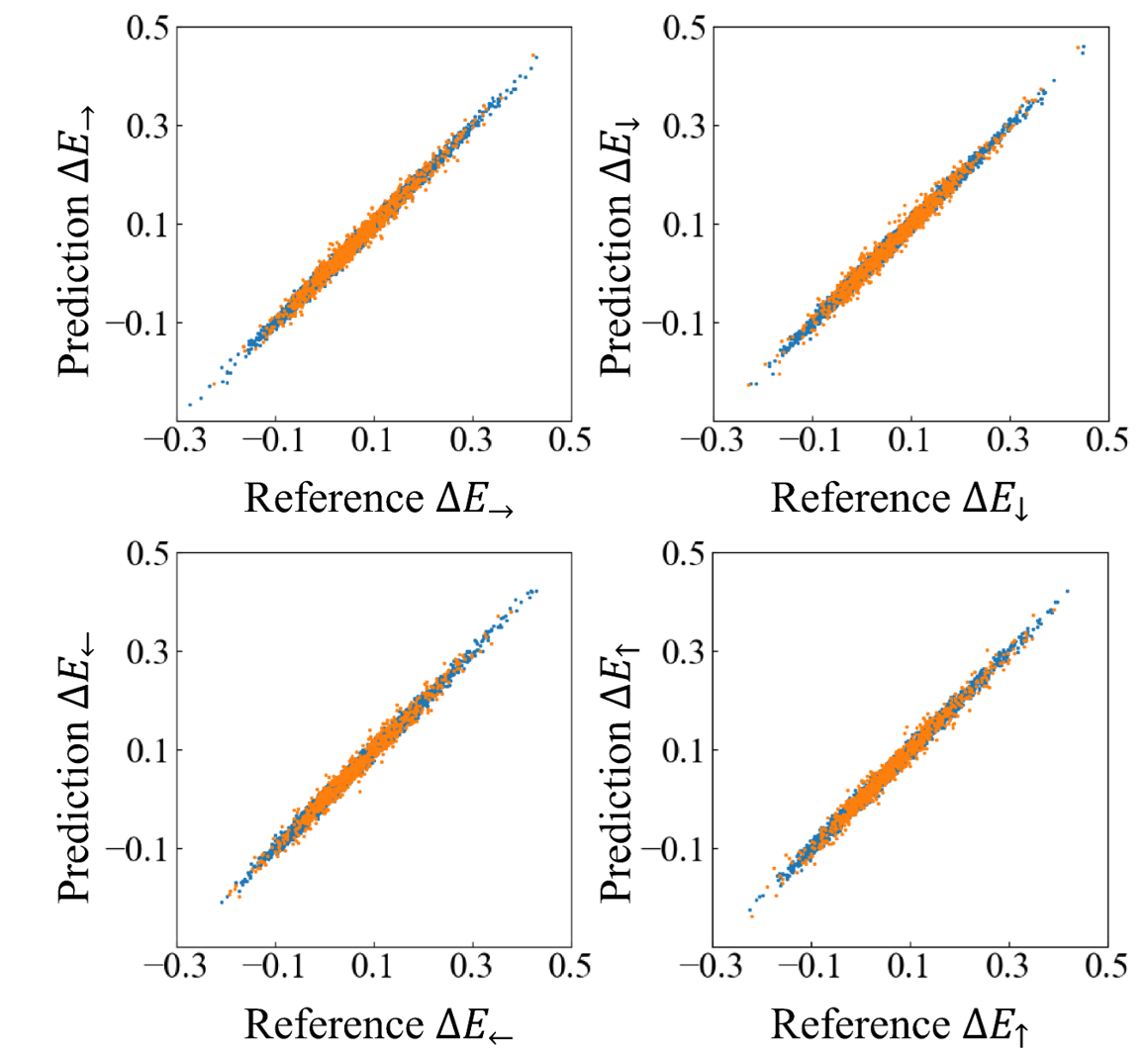}
    \caption{Correlation between the NN predictions and the exact solutions for energy differences $\Delta E_{\rightarrow}$, $\Delta E_{\leftarrow}$, $\Delta E_{\uparrow}$, and $\Delta E_{\downarrow}$, corresponding to the energy cost/gain of hopping to right, left, top, and bottom nearest neighbors of a randomly chosen $f$-electron. The blue dots show the training data prediction with $\text{MSE}=0.041$ and the orange dots show the testing data prediction with $\text{MSE}=0.046$.  }
    \label{fig3}
\end{figure}

The NN model was trained by datasets generated from kMC simulations on a $30\times 30$ square lattice using exact diagonalization method. The following parameters are used: nearest-neighbor hopping $t_{\rm nn}=1$, on-site repulsive energy $U=2\, t_{\rm nn}$, temperature $T=0.05$, $f$-electron density $\rho_{f}=0.187$, and $c$-electron density $\rho_{c}=0.55$. As discussed in the main text, these parameters were chosen in order to realize a low-temperature phase consisting of a mixture of checkerboard patterns and uniform domains that are free of $f$-electrons. 
The training datasets come from both random configurations and ED-kMC simulations for a thermal quench to $T = 0.05$. In average, the NN models for the five different nearest-neighbor configurations shown in Fig.~\ref{fig2} are each trained by at least $10^6$ datasets. 
The Adam optimizer~\cite{adam} with a 0.001 learning rate is used to minimize the loss function, which is defined as the mean square error (MSE) of the energy prediction. We have obtained nice overall agreement between the predicted values and exact calculations, as demonstrated by the validation results shown in Fig.~\ref{fig3}.

\begin{figure}
     \centering
     \includegraphics[width=0.45\textwidth]{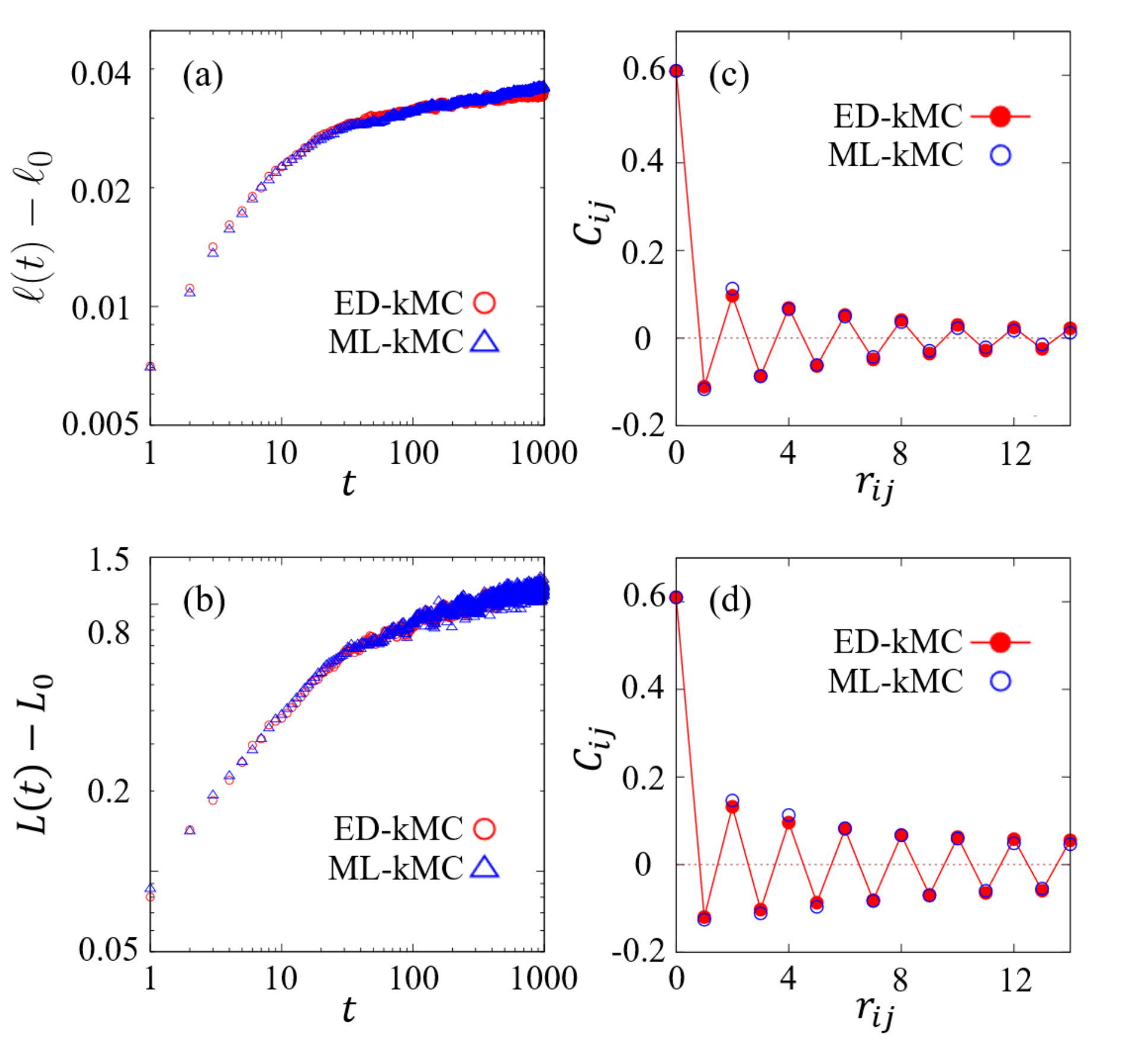}
    \caption{(a) The characteristic length $\ell(t)$ of the $f$-electron configuration $\{n^f_j \}$ as a function of time for ED-kMC and ML-kMC simulations. (b) The time dependence of the characteristic linear size $L$ of the Ising domains associated with the super-clusters.  The panels on the right show the comparison of $f$-electron correlation function $C_{ij} = \langle n^f_i n^f_j \rangle$ after (c) $t=50$ and (d) $t=500$ from the same initial condition.}
    \label{fig4}
\end{figure}

Next, we integrate our NN with the kMC to further benchmark the performance of the ML methods compared with the exact diagonalization (ED) kMC results. We first calculate 100 independent simulations using the ED-kMC with parameters above. Another 100 independent simulations with the same parameters are then carried out by ML-kMC using the trained NN model.  Fig.~\ref{fig4}(a) and (b) show the characteristic lengths of the checkerboard clusters and their super-clusters (or effective Ising domains defined in the main text) obtained from ED and ML-kMC simulations. These two characteristic lengths, $\ell$ and $L$, are obtained from the time-dependent structure factor of $f$-electron configurations and Ising configurations, respectively. In both cases, reasonable agreements between the two methods are obtained. Comparisons of the $f$-electron correlation functions $C_{ij} = \langle n^f_i n^f_j \rangle$ obtained from ED and ML-kMC simulations are shown in Fig.~\ref{fig4}(c) and (d) for 50 and 500 time-steps, respectively, after a thermal quench. The reasonable agreements of the correlation functions indicate the ML-model can successfully capture the relaxation dynamics of the FK model.

\section{Calculation of staggered potential}

\label{sec:stagger-v}

\begin{figure}
     \includegraphics[width=0.45\textwidth]{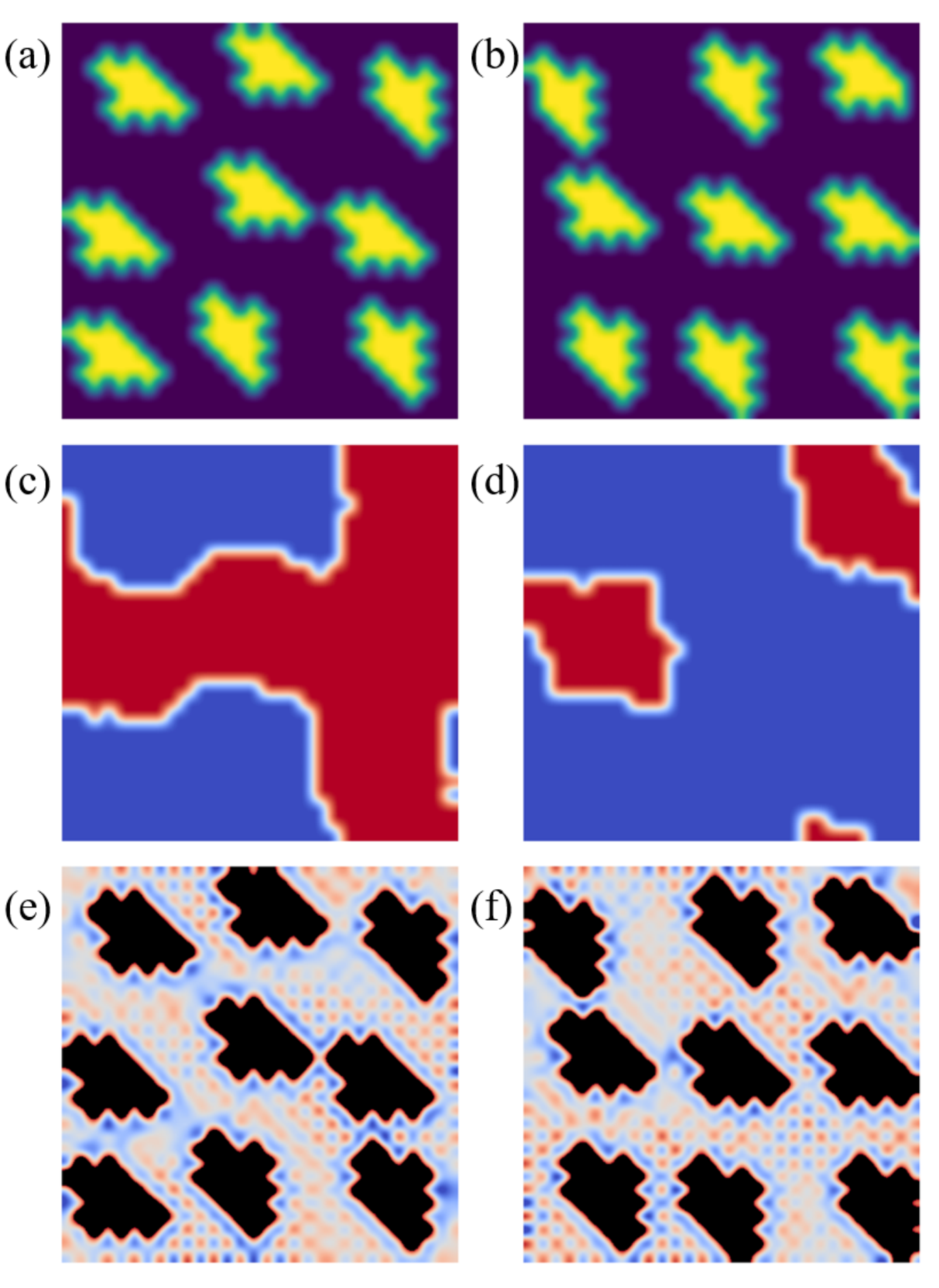}
    \caption{Panels (a) and (b) show two examples of $f$-electron configuration obtained by the block method described in the text. Here yellow regions represent the same-shaped checkerboard cluster that is randomly rotated and placed in each of the 9 blocks; the blue background represents the empty sites. Panels (c) and (d) are the corresponding Ising configuration $\sigma_i$ according to the definition discussed in the main text. Red (blue) regions correspond to $\sigma = +1$ ($-1$). Panels~(e) and (f) show the effective potential experienced by a test $f$-electron created by the collection of checkerboard clusters (shown in black).}
    \label{fig5}
\end{figure}

\begin{figure}[t]
     \includegraphics[width=0.36\textwidth]{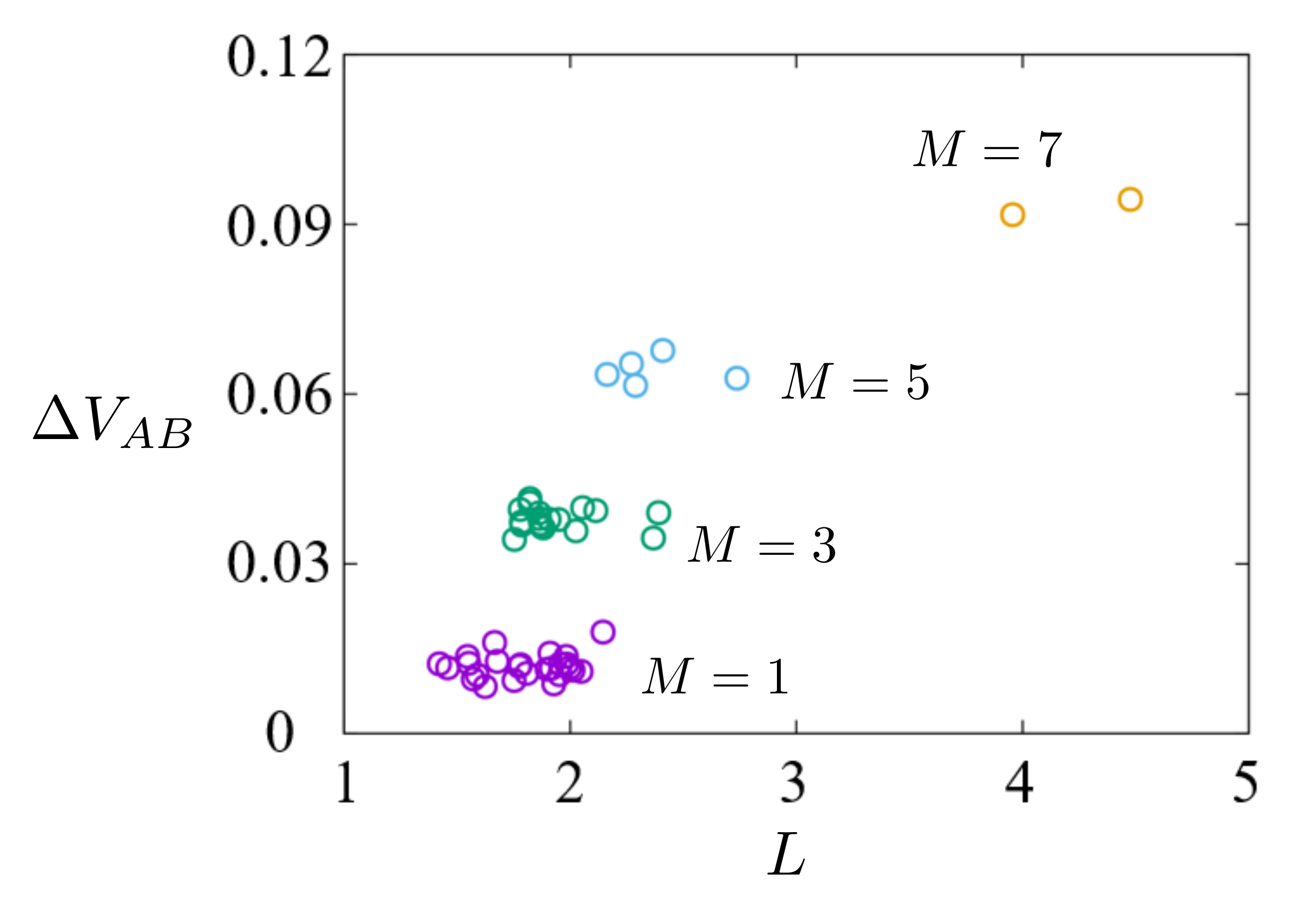}
    \caption{(a) Strength of the staggered potential $\Delta V_{AB} = |V_A - V_B|$ versus the characteristic length $L$ of Ising domains from 50 different $f$-electron configurations generated by the block method. Points of different color share the same block magnetization $M$ discussed in the text.  }
    \label{fig6}
\end{figure}

In this section, we outline the calculation of the staggered potential shown in Fig.~5 in the main text. To investigate the dependence of the staggered potential on the average size of super-clusters, we first generate different super-cluster configurations as follows. A $30\times 30$ lattice with periodic boundary conditions is evenly divided into 9 blocks; see Fig.~\ref{fig5}.  A small checkerboard cluster of arbitrary shape is placed in each block. Different super-cluster configurations can be realized by rotating and/or translating the (same-shaped, same-size) checkerboard cluster within each $10\times10$ block.

For each $f$-electron configuration $\{ n^f_i \}$ generated by the above procedure, we first compute the corresponding Ising configuration $\{\sigma_i\}$; see Fig.~\ref{fig5}(c) and (d) for two examples. Next we compute the resultant characteristic length $L$ of the Ising domains, which represents the typical size of super-clusters, from the structure factor of the Ising variables.  For the same $f$-electron configuration, we also perform Monte Carlo simulations with exact diagonalization to compute the effective potential $V_i=-k_B T\, \text{log} \, \nu_i $ experienced by a test particle, where~$\nu_i$ is the frequency that the test $f$-electron stays at site-$i$. The strength of the staggered potential is defined as $\Delta V_{AB} = \langle \frac{1}{N/2} \bigl| \sum_{i\in {\rm A}} V_i - \sum_{i \in {\rm B}} V_i \bigr| \rangle$. The relation between the staggered potential $\Delta V_{AB}$ and the corresponding Ising domain length $L$ is shown in Fig.~\ref{fig6} for 50 different $f$-electron configurations randomly generated by the block method. 

Interestingly, the data points seem to fall into four groups as highlighted by different colors in Fig.~\ref{fig6}. To understand this unusual result, we first note that one can introduce an Ising variable $\sigma_J$ for the $J$-th block ($J = 1, 2, \cdots, 9$), which is defined as the majority of Ising spins inside this block.  According to the definition  discussed in the main text, the majority of Ising variables in a given block is $\sigma = + 1$ ($-1$) if $f$-electrons of the small checkerboard cluster inside the block resides on A (B) sublattice.  Different $f$-electron configurations generated from the above block procedure can then be classified according to the block Ising spins  $\{ \sigma_J \}$. The total ``magnetization" of block Ising spins is given by $M =  \sum_{J = 1}^9 \sigma_J$. It turns out the four different groups in Fig.~\ref{fig6} correspond to block magnetization $|M| = 1, 3, 5$, and 7. Our results thus indicate that the strength of the staggered potential $\Delta V_{AB}$ depends mainly on the total block magnetization $M$.  Fig.~5 in the main text is then obtained by averaging over data points within each group.

\end{document}